\documentclass[aps,prl,preprint,superscriptaddress]{revtex4-1}
\usepackage{graphicx}

\begin{document}

\title{Quenched Magnon excitations by oxygen sublattice reconstruction in (SrCuO$_2$)$_n$/(SrTiO$_3$)$_2$ superlattices}

\author{M. Dantz}\email[]{marcus.dantz@psi.ch}
\affiliation{Swiss Light Source, Paul Scherrer Institut, CH-5232 Villigen PSI, Switzerland}
\author{J. Pelliciari}
\affiliation{Swiss Light Source, Paul Scherrer Institut, CH-5232 Villigen PSI, Switzerland}
\author{D. Samal}
\thanks{Present adress: Institute of Physics, Sachivalaya Marg, Bhubaneswar-751005, India. }

\affiliation{MESA+ Institute for Nanotechnology, University of Twente, Post Office Box 217, 7500AE Enschede, The Netherlands}
\author{V. Bisogni}
\thanks{Present address: National Synchrotron Light Source II, Brookhaven National Laboratory, Upton, New York 11973, USA. }
\affiliation{Swiss Light Source, Paul Scherrer Institut, CH-5232 Villigen PSI, Switzerland}

\author{Y. Huang}
\affiliation{Swiss Light Source, Paul Scherrer Institut, CH-5232 Villigen PSI, Switzerland}
\author{P. Olalde-Velasco}
\thanks{Present address: Instituto de Ciencias Nucleares, UNAM, Circuito Exterior, Ciudad Universitaria, 04510 Mexico D.F., Mexico}
\affiliation{Swiss Light Source, Paul Scherrer Institut, CH-5232 Villigen PSI, Switzerland}

\author{V. N. Strocov}
\affiliation{Swiss Light Source, Paul Scherrer Institut, CH-5232 Villigen PSI, Switzerland}
\author{G. Koster}
\affiliation{MESA+ Institute for Nanotechnology, University of Twente, Post Office Box 217, 7500AE Enschede, The Netherlands}
\author{T. Schmitt}\email[]{thorsten.schmitt@psi.ch}
\affiliation{Swiss Light Source, Paul Scherrer Institut, CH-5232 Villigen PSI, Switzerland}



\begin{abstract}

The recently discovered structural reconstruction in the cuprate superlattice (SrCuO$_2$)$_n$/(SrTiO$_3$)$_2$ has been investigated across the critical value of $n=5$ using resonant inelastic x-ray scattering (RIXS). We find that at the critical value of $n$, the cuprate layer remains largely in the bulk-like two-dimensional structure with a minority of Cu plaquettes being reconstructed. The partial reconstruction leads to quenching of the magnons starting at the $\Gamma$-point due to the minority plaquettes acting as scattering points. Although comparable in relative abundance, the doped charge impurities in electron-doped cuprate superconductors do not show this quenching of magnetic excitations.

\end{abstract}

\flushbottom
\maketitle
%
%
\thispagestyle{empty}

\section*{Introduction}

Despite the great focus high-Tc cuprates have received in the past 25 years, a comprehensive explanation for superconductivity is still lacking. Since conventional BCS-like phonon mediation has been ruled out, it has been widely accepted that magnetic fluctuations play a significant role. However, their role is still controversial in light of the discovery of competing charge-ordering phenomena \cite{comin_broken_2015, Ghiringhelli_Long-Range_2012, Abbamonte_Spatially_2005,Parker_Fluctuating_2010,li_two-dimensional_2007, Chang_Direct_2012, campi_inhomogeneity_2015}. Therefore, huge effort has recently been put in the investigation of the interplay between structural modifications and magnetic excitations \cite{Dean_Magnetic_2013, Dean_Spin_2012}.

It is thus desirable to investigate the impact of structural changes on the microscopic magnetism of cuprates. Complex oxide thin films and heterostructures in particular provide a multitude of possibilities for artificially tailoring the microscopic interactions at the atomic level since their structure is easily controllable by choice of substrate (i.e. strain) and dimensionality. Recently, developments in the field of epitaxial film growth and lattice engineering have sprouted a manifold of novel materials with interesting properties \cite{samal_manipulating_2015, Mannhart_Oxide_2010, Zubko_Interface_2011}.

Infinite layer SrCuO$_2$ (SCO) in (SrCuO$_2$)$_n$/(SrTiO$_3$)$_2$ superlattices (SL) has been reported to be susceptible to structural changes in the very thin limit due to its polar nature \cite{Zhong_Prediction_2012, Samal_Experimental_2013, Kuiper_Control_2013}. Starting from the infinite layer bulk structure of connected copper-oxygen plaquettes hosting long-range ordered antiferromagnetism, it develops a quasi one-dimensional conformation by re-arranging the oxygen sublattice below a critical thickness of about 5 unit cells (uc), that is caused by a relaxation of the polar electrostatic energy \cite{Samal_Experimental_2013, samal_manipulating_2015}. The  oxygen sublattice rearrangement effectively  causes  the copper-oxygen plaquettes to flip out of the a-b plane (Fig.1(a))  into either the b-c plane (Fig. 1(b)) or the a-c plane (Fig. 1 (c)) resulting in a quasi 1D structure when the SCO thickness is confined below the critical value.

The effect of chemical doping on the magnetic excitations in the cuprates has been studied intensively \cite{le_tacon_intense_2011, dean_persistence_2013, lipscombe_persistence_2007}. However, studies on modifying the collective magnetism by structural modifications, while leaving the stoichiometry intact, are missing so far. In this article, we report on direct measurements of the influence of a structural reconstruction on collective magnetic excitations in cuprate 2D layers. While being completely reoriented below the reconstruction threshold thickness, the SLs around the critical thickness exhibit an instability towards flipping, i.e. while the majority of plaquettes are still in bulk like configuration, a minority of plaquettes is already flipped, forming impurities in the 2D layers. These impurities act as scattering points for magnons, quenching them around the $\Gamma$-point. This allows us to compare the present case of dimensionality-induced defect scattering to the known effects of chemical doping in the cuprates.

We employ resonant inelastic x-ray scattering (RIXS), which has been established as a powerful tool for the simultaneous investigation of structural, electronic and magnetic properties especially for cuprates \cite{Monney_Determining_2013, Dean_Spin_2012, Schlappa_Collective_2009, Lee_Role_2013}. RIXS is sensitive to orbital and ligand-field excitations \cite{Schlappa_Spin-orbital_2012, Sala_Energy_2011, Bisogni_Orbital_2015}, as well as to elementary magnetic excitations, such as the evolution from magnon to paramagnon character upon doping \cite{Dean_High-Energy_2013,  le_tacon_intense_2011, Zhou_Persistent_2013, Guarise_Anisotropic_2014}. The experiment was performed with the RIXS spectrometer of the ADRESS beamline of the Swiss Light Source at the Paul Scherrer Institut \cite{strocov_high-resolution_2010, Ghiringhelli_SAXES_2006}, which allows for a combined energy resolution better than 130 meV. A complete description of the experimental conditions can be found in the supplementary information.

\begin{figure}[tb]
\centering
\includegraphics[width=\linewidth]{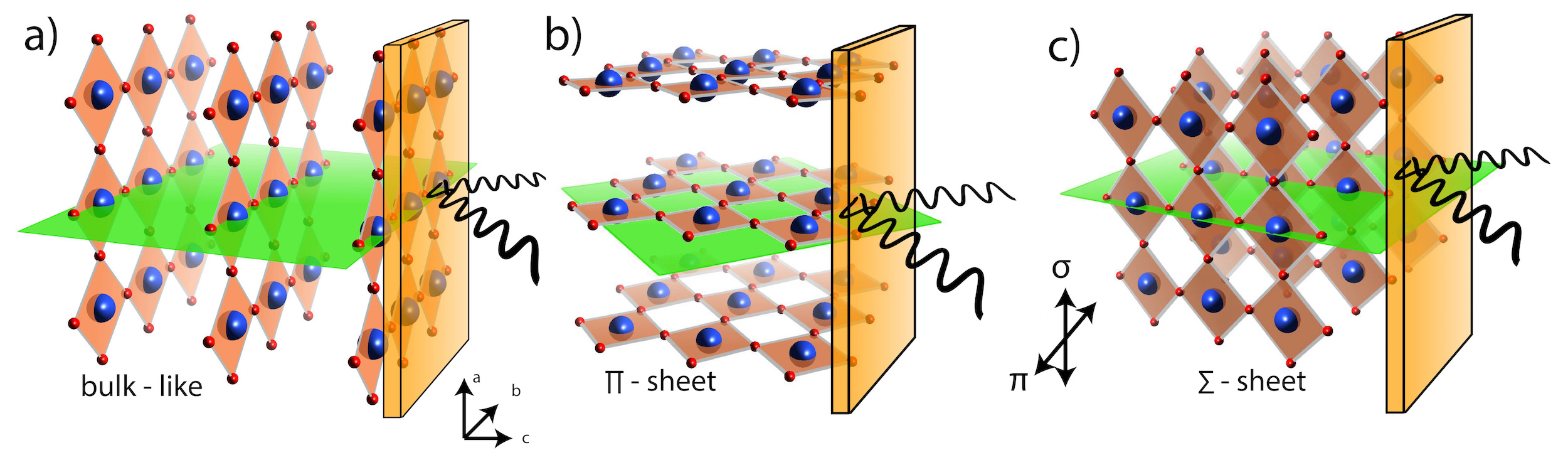}
\caption{\label{}Sketches of the structure with respect to the scattering plane (green plane), for (a) the bulk like infinite layer, (b) the $\Pi$ sheet orientation, (c) the $\Sigma$-sheet orientation. Red (blue) balls depict oxygen (copper). Strontium atoms are omitted for clarity.}
 \end{figure}

\section*{Results}
\subsection*{Structure}
In Fig. 1 we show the different structural orientations possible in these superlattices. Fig 1(a) depicts the bulk-like structure with the copper-oxygen plaquettes oriented in the a-b plane forming a quasi 2D checkerboard-like structure, i.e parallel to the sample surface, thus being perpendicular to the growth direction. When reducing the thickness below the reorientation threshold of 5 uc, the orientation of the oxygen sublattice changes by repositioning some of the oxygen ions from the CuO$_2$ plane to the Sr plane (i.e. the b-c plane), thus effectively flipping the copper-oxygen plaquettes along one axis perpendicular to the c axis. Since the a and b crystal directions are twinned, the flipping occurs with equal probability along both directions, creating domains of the two different orientations (see Fig. S1 in the supplementary information), as depicted in Fig. 1(b) and (c). As the copper plaquettes are now oriented either in the a-c or b-c plane, they are not continuous along c-direction due to the limited thickness of SCO layer, effectively reducing them to a quasi one-dimensional structure of width $n$. Note that structurally, these two orientations are equivalent, but our scattering geometry introduces a distinction between them as one of the two possible orientations is lying in the scattering plane (Fig. 1(b)) and one is oriented perpendicular to it (Fig. 1(c)). Therefore, we introduce the nomenclature of calling the former $\Pi$ sheets and the latter $\Sigma$ sheets, following the notation of light polarisation in the respective direction.

\begin{figure}[tb]
\centering
\includegraphics[width=\linewidth]{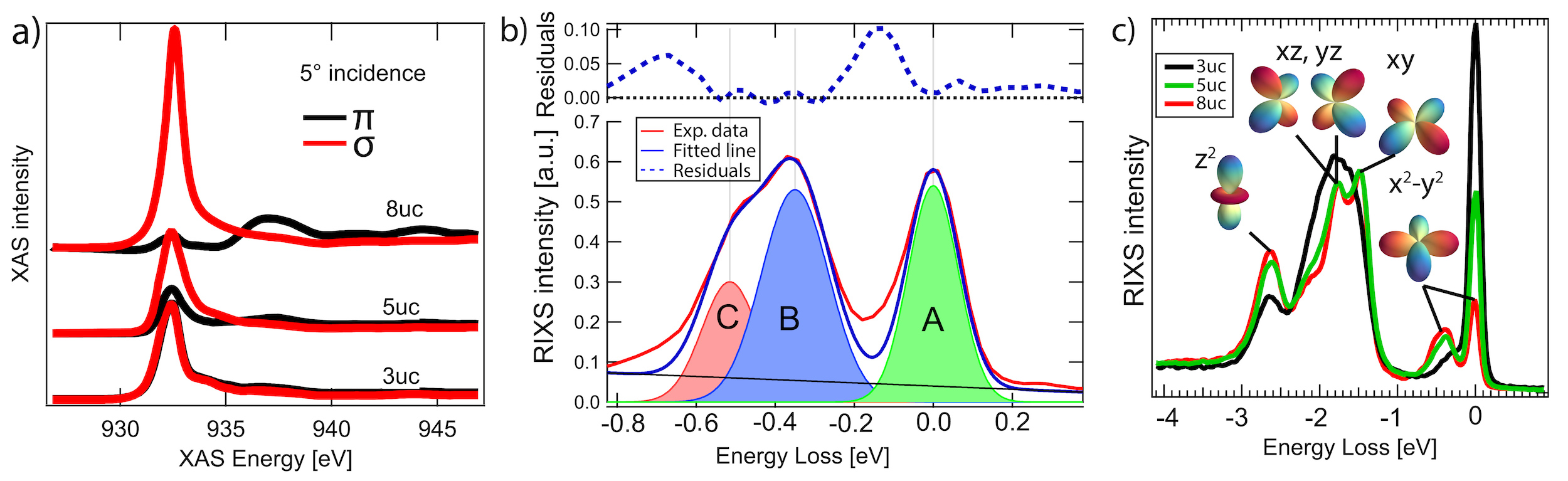}
 \caption{\label{}a) Cu L$_3$ X-ray absorption spectra for the different superlattices measured, showing the successively vanishing linear dichroism. The XAS of the 80uc thick film is virtually identical to the 8uc SL and therefore left out. b) Exemplary fit of the magnetic part. The remaining spectral weight between the elastic line (A) and the magnon (B) can be attributed to phonons including higher order harmonics \cite{Lee_Role_2013}. The Multi Magnon part (C) has been fitted with a Gaussian line, as the higher order contributions are negligible for the position of the single magnon peak. c) Cu L$_3$ RIXS data for the superlattices. Note the remaining $d_{z^2}$ peak at 2.7eV, which is persisting across the atomic reorientation threshold (see text). Both b) and c) are taken at 10K with $\sigma$ polarization at $q=0.45 (r.l.u.)$.}
 \end{figure}

In bulk single crystals, SrCuO$_2$ forms quasi one-dimensional chains. However, it has also been reported that if grown under pressure and high temperature, a 2D planar structure can also be realised. \cite{Zaliznyak_Spinons_2004, Walters_Effect_2009, takano_acuo2_1989} In particular, if grown on SrTiO$_3$, SrCuO$_2$ grows in a two-dimensional structure, as can be inferred from the structural data in Ref. \cite{Samal_Experimental_2013}, as well as from a polarization analysis of the XAS data in Fig. 2a, in which we show Cu L$_3$ x-ray absorption spectra (XAS), taken at an incidence angle of $5$ degrees for 3, 5, and 8uc SL. The XAS data shows a vanishing excitonic peak at 931eV for $\pi$ polarization, whereas for $\sigma$ polarization, this feature is dominating the spectrum. This is incompatible with the 1D crystal structure as found in bulk crystalline SrCuO$_2$ as it proves that the unoccupied d$_{x^2-y^2}$ orbital lies parallel to the surface of the sample. In the 1D bulk crystalline SrCuO$_2$, the unoccupied orbital lies perpendicular to the surface \cite{Hyatt_High-pressure_2004}, which would produce either no dichroism (in the case of the chains running perpendicular to the scattering plane), or it would give a strong excitonic peak in the $\pi$ polarization and a vanishing signal in the $\sigma$ polarization otherwise. When tuning the thickness of the films measured in this work, we see the footprint of the change in crystal structure in the evolution of the XAS in Fig. 2a, exactly as expected from this geometrical argument. The disappearance of the dichroism upon crossing the structural reorientation threshold can be explained by the copper-oxygen plaquette sheet reorientation, shown in Fig. 1b and 1c, assuming an equal distribution of sheet orientations between $\Pi$ and $\Sigma$ sheets and an average domain size much smaller than the x-ray beam spot (in this case ca. $4$x$50\mu m^2$). As local tetragonal symmetry is preserved, both sheet orientations have the hole in the $d_{x^2-y^2}$ orbital lying in the sheet plane, thus being only accessible with $\sigma$ polarised light for the $\Sigma$ sheets, and $\pi$ polarised light for the $\Pi$ sheet.  This dichroism emphasises that above the reorientation thickness threshold, this compound indeed forms a 2D structure, similar to CaCuO$_2$ investigated in Ref. \cite{Minola_Magnetic_2012} Furthermore, this results in the vanishing dichroism observed when crossing the reorientation threshold. 

\subsection*{Orbital structure}

It has been observed that the Cu L$_3$ XAS data resembled an orbital reconstruction, but a clear interpretation was still lacking \cite{Samal_Experimental_2013}. An equal occupation of the $d_{x^2-y^2}$ and the $d_{z^2}$ orbitals was assumed from vanishing linear dichroism observed in XAS. This would, however, be incompatible with local tetragonal symmetry as these two orbitals would have to be degenerate. Therefore, direct access to the energy levels of the different d-orbitals is desirable. In Fig. 2c, we show Cu L$_3$ RIXS spectra for 3, 5 and 8uc superlattices containing the spectral footprint of the different dd-excitations, i.e. transitions of the electron vacancy from the $d_{x^2-y^2}$ orbital to other d-orbitals. The persistence of the $d_{z^2}$ peak upon crossing the reorientation threshold is a fingerprint of the persistence of local tetragonal symmetry, discouraging the picture of orbital reconstruction. From the RIXS data, we can immediately extract the crystal field splitting parameters for the bulk-like cases. Using the method presented in Ref. \cite{Sala_Energy_2011}, we obtain values of $D_s = 0.429 \pm 0.005$ eV, $D_t = 0.197 \pm 0.005$ eV and $10D_q = 1.486 \pm 0.003$ eV, which are similar to the values obtained for bulk CaCuO$_2$ \cite{Sala_Energy_2011}. Details are shown in the Supplementary Information.

\subsection*{Magnetic excitations}
Next, we investigate the magnetic excitations in these superlattices. RIXS is an ideal tool for investigating magnetic excitations in thin films as it is sensitive even to layers of one unit cell thickness \cite{Minola_Magnetic_2012,Dean_Spin_2012}. A sketch of the experimental setup can be found in the supplementary materials, Fig. S3. In Fig 3, we show momentum dependent RIXS measurements with incident polarization chosen in order to enhance single magnon excitations \cite{Ament_Theoretical_2009}, i.e. $\sigma$ polarisation in grazing incidence geometry (corresponding to negative $q$-transfer values) and $\pi$ polarisation in grazing exit geometry (corresponding to positive $q$-transfer values) for the 80 unit cell thick film (Fig. 3(a)), the 8 and 5uc SL (Fig. 3(b) and (c), respectively). For the 3 uc SL however, due to the change in geometry, we restrict ourselves to $\pi$ polarisation in order to selectively excite the $\Pi$ sheets. This choice was made because the $\Sigma$ sheets are continous only perpendicular to the scattering plane. In order to obtain the exact positions of the magnetic excitations, we performed a multi peak fit as shown in Fig. 2c. Note that single magnon excitations (Peak B) is close to the resolution limit as it is very similar in width to the elastic scattering line (Peak A) and can thus be identified as a coherent magnon excitation. Peak C is an excitation of multiple magnons at once with highest contribution of a bigmanon. Note that a Bimagnon peak does not necessarily have twice the energy of a single magnon excitation \cite{braicovich_magnetic_2010}. The higher order contributions have been neglected as can be seen from the small high energy tail as they do not play a role for the exact location of the single magnon excitation. For the 80uc thick film as well as the 8uc and 5uc SL, we observe a clear magnon dispersion with a periodicity of the lower bound of $\pi$. Note that we use the convention commonly used in RIXS of defining the momentum transfer so that the antiferromagnetic ordering vector is at (h,k,l)=(1, 1, 0)$ \pi$. For a 1D compound such as bulk single crystal SrCuO$_2$, a spinon dispersion with a lower bound periodicity of $0.5 \pi$ would be expected, as has been observed by some of us in the related compound Sr$_2$CuO$_3$ in Ref. \cite{Schlappa_Spin-orbital_2012} and by inelastic neutron scattering in both SrCuO$_2$ and Sr$_2$CuO$_3$ \cite{Walters_Effect_2009, Zaliznyak_Spinons_2004}. We observe that the magnetic spectral weight of both the 80uc thick film and the 8uc SL have a markedly different behaviour compared to the magnetic excitation in the aforementioned references. This is further evidenced by the magnon behaviour of the magnetic excitations in the $\Gamma -$ M  direction of the 8uc sample shown in the supplementary information, Fig. S4. Both the 80uc and the 8uc cases are reasonably similar, while for the 5uc SL, we observe a reduction of the spectral weight of the magnetic excitation starting around the $\Gamma$ point. For the 3uc sample, the magnetic excitations seem to be quenched completely. This can also be seen in the constant momentum cuts shown in Fig. 4(a). For the 5 uc SL, the spin excitations show a more sophisticated behaviour. While virtually absent in the lowest momentum cut, the spin excitations close to the AF zone boundary are almost as strong as for the bulk sample. We also observe that multi-magnon excitations are unusually strong compared to other undoped cuprates \cite{braicovich_dispersion_2009}. The relative strength of multi-magnon excitations has been linked to increased quantum fluctuations \cite{Dean_Spin_2012}, which might be caused by strain induced by the substrate or due to the reduced dimensionality.

\begin{figure}[tb]
\centering
\includegraphics[width=\linewidth]{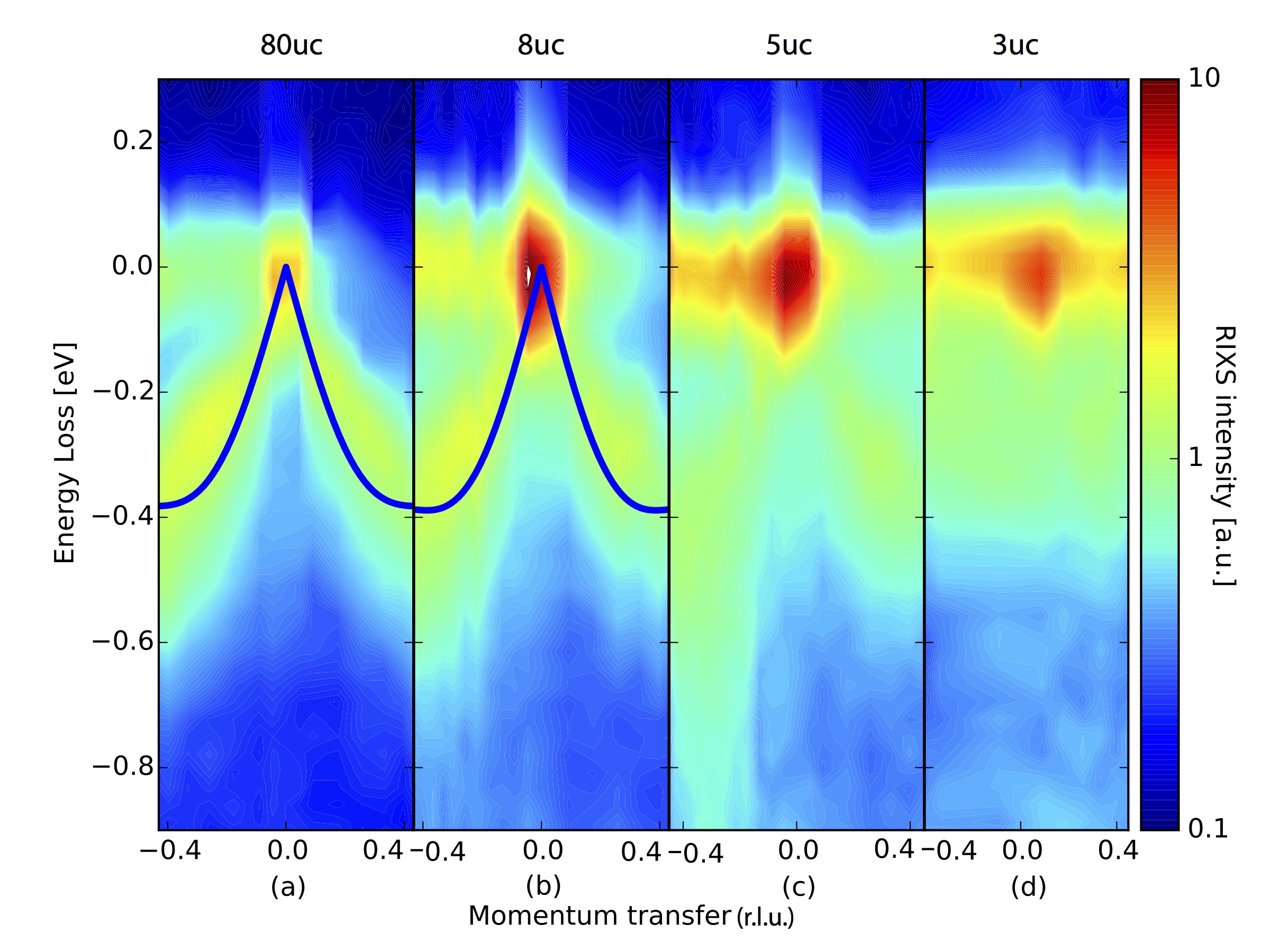}
 \caption{\label{}Contour plots of the RIXS data for (a) the 80 uc thick film, (b) the 8uc superlattice, (c) the 5uc superlattice and (d) the $\Pi$ sheet in the 3uc superlattice. The solid lines in (a) and (b) are the fits to linear spin wave theory. Note for the 8 and 5uc superlattices the increasing leakage of spectral weight from the coherent magnon excitation to the incoherent region at higher energies. See text for detailed experimental conditons.}
 \end{figure}

\begin{figure}[tb]
\centering
\includegraphics[width=\linewidth]{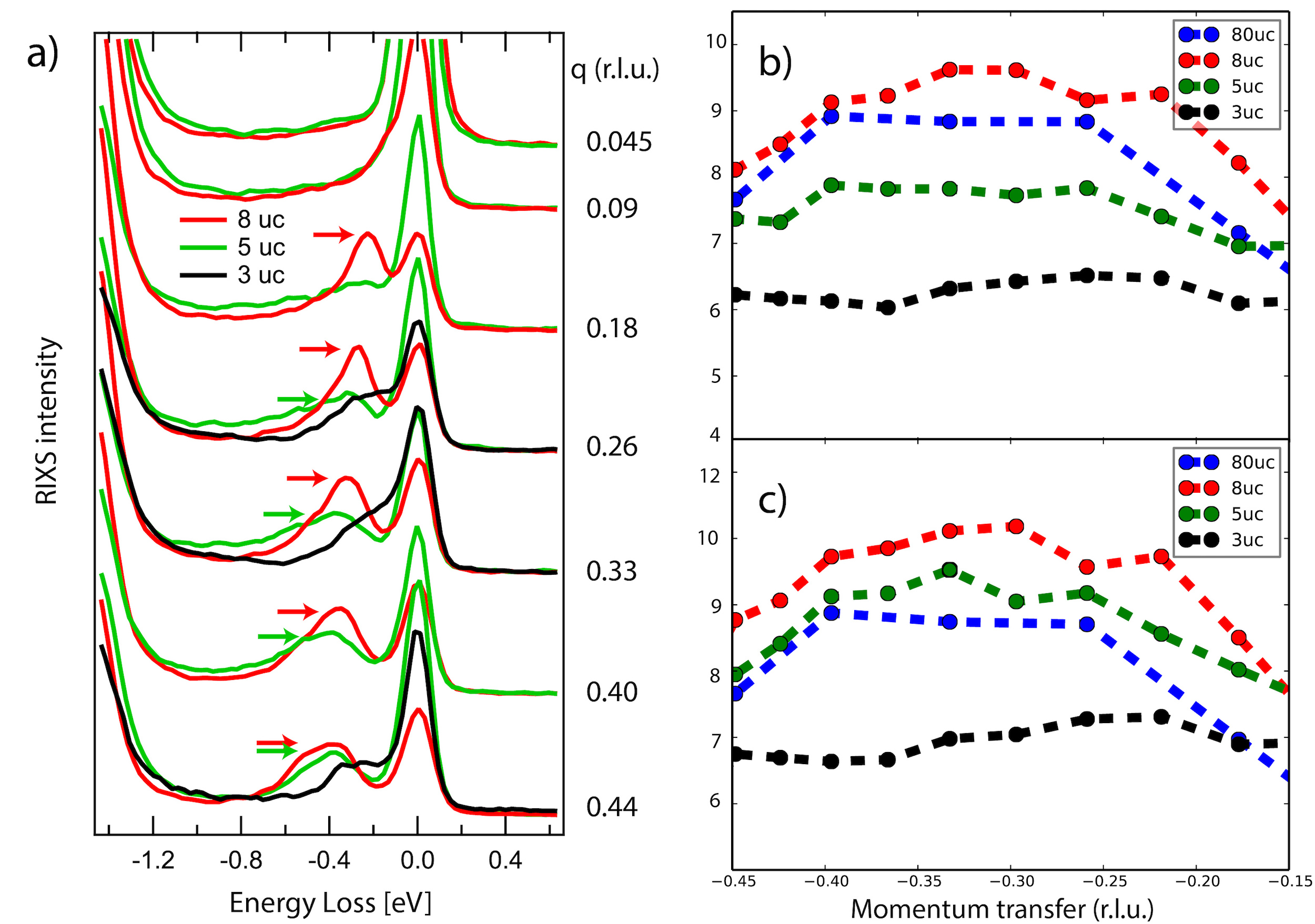}
 \caption{\label{}(a) Constant momentum cuts for the 3, 5 and 8uc cases, showing the quenching of the magnetic spectral weight. The arrows denote the maximum intensity of the low energy magnetic excitation. (b and c) Integrated spectral weight of the magnetic excitations, for (b) the low energy part (up to 0.6eV), (c) the whole magnetic regime up to 0.9eV. Note that for the 80uc, 8uc and 3uc the ratio of spectral weight remains constant in (b) and (c), while the integrated signal of the 5uc cases increases significantly (Note that all spectra are for $\sigma$ polarisation, i.e probing the $\Sigma$ sheet for the 3 uc film). It should be noted that due to the different structual conformation, the 3uc can only act as a rough reference.}
 \end{figure}

Although quenched for low $q$, the magnetic excitations in the 5uc SL are located at similar energies as the 8uc SL for higher q, see Fig. 3. Spectral weight is transferred from the low energy part at about $300$ meV to higher energies, as can be seen in Fig. 3(c). For the 3 uc SL, the low energy part of the magnetic spectral weight is quenched across the entire BZ, see Fig. 3(d). In Fig. 4(a), we show cuts of the contour plots for different $q$ and different SCO thicknesses. We observe that the difference in spectral weight between the 5uc and 8uc cases is largest around the $\Gamma$ point and is reduced towards the $X$ point. Furthermore, for low $q$ transfer, we see a pronounced high energy tail between $0.4$ eV and $0.9$ eV, which can also be observed in Fig. 3(c). However, the spectral weight of the low energy part of the magnetic excitation in the 5uc case is almost matching that of the bulk-like 8uc one for large $q$, see Fig. 4a. We interpret the low energy part as a coherent magnon excitation and the higher energy part as an incoherent excitation. In order to investigate this shift of spectral weigh, we sum up the coherent part of the magnetic excitations (0.1eV to 0.6 eV), shown in Fig. 4b, and the both the coherent and incoherent part (0.1 eV to 0.9 eV) in Fig. 4c. We can see that the 8uc and 5uc films are castly different when taking only the coherent part into account, while being almsot identical when taking the whole range of magnetic excitations into account. However, we can observe slight differences between the different samples, even between the 8uc and the reference thick film of 80uc. We attribute this to small interface effects, as have been demonstrated in Refs \cite{Dean_Spin_2012,Minola_Magnetic_2012}. For the 3uc sample, we see still some amount of spectral wight, which can be explained by spin excitations of incoherent nature throughout the Brillouin zone.

From the dispersion of the magnon excitations of the 8 uc and 80 uc cases it is possible to extract the superexchange constant $J$ by using a linear spin wave approach \cite{Coldea_Spin_2001, headings_anomalous_2010, Minola_Magnetic_2012}. For thinner SCO layers ($5$ and $3$ uc SLs), the quenching of the magnetic spectral weight around the $\Gamma$ point prohibits such an analysis. As a starting point for our fitting, we used Ref. \cite{headings_anomalous_2010} which employed INS to measure the magnetic dispersion of a La$_2$CuO$_4$ bulk crystal, and extracted $J=143$ meV, $J_c=-58$meV and $J^{\prime\prime} = J_c^{\prime}= 2$ meV. Taking into account our resolution of $130$ meV, we assume the latter two to be equal to 0. The measurements were carried out in the $\Gamma-X$ direction, therefore $k_y=0$. The linear spin wave equation used here is given as Eq. 1 in the Methods section. The linear spin wave fit results are shown in Fig. 3(a) and (b) on top of the contour plots as solid lines. The difference between the 80uc film and 8uc SL are within error bars, confirming that the 8uc case is also a good representative of the unperturbed bulk-like system in terms of the superexchange J. The fitted parameters for the 8uc heterostructure are $J_8=163\pm6$ meV, $J_{c8}=-52\pm6 $ meV and for the 80uc thick film we obtain $J_{80}=162 \pm6$ meV, $J_{c80} = -52\pm6$ meV. This is in reasonable agreement with the values for CCO (which shares a similar structure), for which $J=157$ meV and $J_c=-49$ meV were extracted \cite{Minola_Magnetic_2012}. 

When comparing the absorption measurements (Fig. 2a), we find that in the 5 uc SL there is a remanent linear dichroism, although smaller than that of the 8uc SL. We attribute this to a structure of largely unperturbed Cu-O plaquettes, which is interrupted by a minority of flipped plaquettes. These minority plaquettes act as scattering points for magnons, breaking the antiferromagnetic long-range order, creating finite size effects. This allows us to estimate the average distance between two flipped plaquettes. We take the $q$-value for which the coherent magnetic peak is about half as intense as for the bulk-like sample, which is about $q=0.26 (r.l.u.) = 2\pi/\lambda$, where $\lambda$ is the wavelength of the magnon which we take as the minimum coherence length needed. This yields $\lambda = a / 0.26 \approx 4a$, i.e. an average distance between flipped plaquettes of 4 lattice constants.

\section*{Discussion}
The linear dichroism observed in the XAS spectra as well as the magnetic dispersion are clear indications that the bulk-like SLs and the thick film are indeed in a 2D structure. For the thinner films, a contribution of reoriented plaquettes make the structure less obvious. The quenching of spectral weight for the 5uc film near the $\Gamma$ point and the overall disappearance of a coherent magnetic excitation for the 3uc film show that the reorientation is gradual over several thicknesses, which is consistent with the evolution of the c-axis parameter measured in Ref \cite{Samal_Experimental_2013}.

Our conclusion is that in the 5uc SL the majority of copper-oxygen plaquettes are still in the bulk-like planar configuration while the minority of the plaquettes are creating scattering points for magnons.  Therefore, we conclude that the transfer of spectral weight to higher loss energies for low and intermediate $q$ (which corresponds to long wavelength magnons) is the result of local spin-flip excitations that cannot form coherent magnons, while for large $q$ (i.e. short-wavelength magnons) the finite size effect induced by the flipped plaquettes is less influential. In order to estimate the amount of flipped plaquettes in the 5 uc film, we can estimate an upper boundary by integrating the coherent magnetic spectral weight. We compare the magnetic spectral weight of the 5uc and 8uc samles at highest q. Therefore, we can assume that even in the 5uc sample no incoherent scattering takes place. Under this assumption, the difference in spectral weight is only due to the flipped plaquettes in the 5uc film and we calculate an upper limit of $8\pm2\%$ of flipped plaquettes.

This loss of magnetic coherence due to finite size effects is of considerable interest in the context of high-temperature superconductors. For hole-doped cuprates, the doped positive charge has dominantly Zhang-Rice exciton character and is thus mainly located on the oxygen atoms surrounding the Cu ion \cite{Monney_Determining_2013, Zhang_Effective_1988}, while for electron doped cuprates the doped charge is donated by the donor ion directly to the Cu ion, leading to a Cu d$^{10}$ (i.e. S=0) impurity \cite{Armitage_Progress_2010}. In a simple approximation, the doped charge in electron-doped cuprates can be seen as an impurity in the otherwise antiferromagnetically correlated Cu ions, comparable to the impurity represented in the present case by the flipped plaquettes.

Previous RIXS experiments on electron cuprates, e.g. Nd$_{2-x}$Ce$_x$CuO$_4$ (NCCO) were carried out up to a doping level of $x=0.18$ \cite{Lee_Asymmetry_2014, Ishii_High-Energy_2014}. This corresponds to an average distance of impurities of $2/0.18 \approx 11$ Cu sites, which corresponds to a $q$-value of 0.09 $(r.l.u.)$. While this distance is larger by a factor of 2-3 compared to the average distance of impurities in the SCO/STO superlattices investigated here, the paramagnos should be quenched at the corresponding $q$-value of $q= 0.09 (r.l.u.))$ in the electron doped cuprates. However, Fig. 3 of Ref. \cite{Ishii_High-Energy_2014} clearly shows they are not. Therefore, we conclude that the scattering mechanism responsible for the quenching of the spectral weight in the present case is inherently different from that found in the superconducting cuprates and implies that the breaking of antiferromagnetic order found in the latter is not static.

In conclusion, we have performed RIXS and XAS at the Copper L$_3$ edge in order to unravel the interplay between local structure and magnetic excitations of SCO layers at the unit cell level in the (SrCuO$_2$)$_n$/(SrTiO$_3$)$_2$  superlattices. We found an antiferromagnetic nearest neighbour superexchange constant of $162\pm6$ meV for the 80uc film and $163\pm6$ meV for the 8uc superlattice, very close to the value obtained for infinite layer CCO. We further observed that around the structural reconstruction threshold of 5uc the majority of plaquettes are still in the bulk like planar structure while a minority is already reoriented. These reoriented plaquettes form scattering points for magnons, due to which magnon spectral weight around the $\Gamma$ point is significantly reduced. For the 3uc case, no coherent magnons have been observed. This mechanism of impurity scattering is not present in the comparable case of electron-overdoped cuprate superconductors, thus pointing towards an influence of itinerancy on the interplay of magnetic and electronic properties in the cuprate superconductors. Our findings further emphasise the ability of this class of superlattices to act as an easily modifiable model system of 2-dimensional layered cuprates.


\section*{Methods}

\subsection*{Experiment}
RIXS spetra were taken at the maximum of the Cu L$_3$ edge (931.1 eV), i.e. at the $2p_{3/2}$ to $3d$ resonance. A scattering angle of $130$ degree was used for all spectra with incidence angles varying between 5 and 125 degrees with respect to the sample surface. The samples were oriented ex-situ and measured at a pressure of $5\cdot 10^{-10}$ mbar at a temperature of 10K. The acquisition time for each spectrum varied between 60 and 180 minutes for different sample geometries. The data was normalized to the full spectral weight of the crystal field excitations. The momentum transfer $q$ is calculated by the projection of the absolute momentum transfer in the two-dimensional or quasi one-dimensional sample orientation, see the sketch in the supplementary information Fig. S3.

\subsection*{Linear Spin wave theory}
For the extraction of the magnetic exchange parameter $J$, we used a linear spin wave approach in line with previous work \cite{headings_anomalous_2010}. The full equation for the linear spin wave theory is 

\begin{equation}
E_q=2Z_c \sqrt{A_q^2-B_q^2}
\end{equation}

with $A_q=J-J_c/2-(J^{\prime}-J_c/4)(1-\nu_h\nu_k)-J^{\prime\prime}[1-(\nu_{2h}+\nu_{2k})/2]$, $B_q=(J-J_c/2)(\nu_k+\nu_h)/2$ and $\nu_x=cos(2\pi x)$. Since we are only looking at $\Gamma$ - $X$ direction, it follows that $\nu_h = 1$ and since we set $J^{\prime}=J^{\prime\prime}=0$, Eq. 1 reduces to

\begin{equation}
E_q = 2Z_c\sqrt{(J-Jc/2+Jc/4(1-cos(2\pi h)))^2{-((J-Jc/2)(cos(2\pi h)+1)/2)^2}} .
\end{equation}
$Z_c$ has been determined to be 1.18. \cite{Coldea_Spin_2001}

\section*{Acknowledgements}

This research is funded by the Swiss National Science Foundation and its Sinergia network Mott Physics Beyond the Heisenberg (MPBH) model. Experiments have been performed at the ADRESS beamline of the Swiss Light Source at the Paul Scherrer Institut. J.P. and T.S. acknowledge financial support through the Dysenos AG by Kabelwerke Brugg AG Holding, Fachhochschule Nordwestschweiz, and the Paul Scherrer Institut. This research has been partially funded by the Swiss National Science Foundation within the D-A-CH programme (SNSF Research Grant 200021L 141325). V. B. and P.O. acknowledge financial support from the European Communitys Seventh Framework Programme (FP7/2007-2013) under Grant Agreement No. 290605 (PSIFELLOW/COFUND).






\begin{thebibliography}{10}
\expandafter\ifx\csname url\endcsname\relax
  \def\url#1{\texttt{#1}}\fi
\expandafter\ifx\csname urlprefix\endcsname\relax\def\urlprefix{URL }\fi
\providecommand{\bibinfo}[2]{#2}
\providecommand{\eprint}[2][]{\url{#2}}

\bibitem{comin_broken_2015}
\bibinfo{author}{Comin, R.} \emph{et~al.}
\newblock \bibinfo{title}{Broken translational and rotational symmetry via
  charge stripe order in underdoped {YBa}2cu3o6+y}.
\newblock \emph{\bibinfo{journal}{Science}} \textbf{\bibinfo{volume}{347}},
  \bibinfo{pages}{1335--1339} (\bibinfo{year}{2015}).

\bibitem{Ghiringhelli_Long-Range_2012}
\bibinfo{author}{Ghiringhelli, G.} \emph{et~al.}
\newblock \bibinfo{title}{Long-{Range} {Incommensurate} {Charge} {Fluctuations}
  in ({Y},{Nd}){Ba}2cu3o6+x}.
\newblock \emph{\bibinfo{journal}{Science}} \textbf{\bibinfo{volume}{337}},
  \bibinfo{pages}{821--825} (\bibinfo{year}{2012}).

\bibitem{Abbamonte_Spatially_2005}
\bibinfo{author}{Abbamonte, P.} \emph{et~al.}
\newblock \bibinfo{title}{Spatially modulated '{Mottness}' in
  {La}2-{xBaxCuO}4}.
\newblock \emph{\bibinfo{journal}{Nat Phys}} \textbf{\bibinfo{volume}{1}},
  \bibinfo{pages}{155--158} (\bibinfo{year}{2005}).

\bibitem{Parker_Fluctuating_2010}
\bibinfo{author}{Parker, C.~V.} \emph{et~al.}
\newblock \bibinfo{title}{Fluctuating stripes at the onset of the pseudogap in
  the high-{Tc} superconductor {Bi}2sr2cacu2o8+x}.
\newblock \emph{\bibinfo{journal}{Nature}} \textbf{\bibinfo{volume}{468}},
  \bibinfo{pages}{677--680} (\bibinfo{year}{2010}).

\bibitem{li_two-dimensional_2007}
\bibinfo{author}{Li, Q.}, \bibinfo{author}{H\"ucker, M.}, \bibinfo{author}{Gu,
  G.~D.}, \bibinfo{author}{Tsvelik, A.~M.} \& \bibinfo{author}{Tranquada,
  J.~M.}
\newblock \bibinfo{title}{Two-{Dimensional} {Superconducting} {Fluctuations} in
  {Stripe}-{Ordered} {La}1.875ba0.125cuo4}.
\newblock \emph{\bibinfo{journal}{Phys. Rev. Lett.}}
  \textbf{\bibinfo{volume}{99}}, \bibinfo{pages}{067001}
  (\bibinfo{year}{2007}).

\bibitem{Chang_Direct_2012}
\bibinfo{author}{Chang, J.} \emph{et~al.}
\newblock \bibinfo{title}{Direct observation of competition between
  superconductivity and charge density wave order in {YBa}2cu3o6.67}.
\newblock \emph{\bibinfo{journal}{Nat Phys}} \textbf{\bibinfo{volume}{8}},
  \bibinfo{pages}{871--876} (\bibinfo{year}{2012}).

\bibitem{campi_inhomogeneity_2015}
\bibinfo{author}{Campi, G.} \emph{et~al.}
\newblock \bibinfo{title}{Inhomogeneity of charge-density-wave order and
  quenched disorder in a high-{Tc} superconductor}.
\newblock \emph{\bibinfo{journal}{Nature}} \textbf{\bibinfo{volume}{525}},
  \bibinfo{pages}{359--362} (\bibinfo{year}{2015}).

\bibitem{Dean_Magnetic_2013}
\bibinfo{author}{Dean, M. P.~M.} \emph{et~al.}
\newblock \bibinfo{title}{Magnetic excitations in stripe-ordered
  {La}1.875ba0.125cuo4 studied using resonant inelastic x-ray scattering}.
\newblock \emph{\bibinfo{journal}{Phys. Rev. B}} \textbf{\bibinfo{volume}{88}},
  \bibinfo{pages}{020403} (\bibinfo{year}{2013}).

\bibitem{Dean_Spin_2012}
\bibinfo{author}{Dean, M. P.~M.} \emph{et~al.}
\newblock \bibinfo{title}{Spin excitations in a single {La}2cuo4 layer}.
\newblock \emph{\bibinfo{journal}{Nat Mater}} \textbf{\bibinfo{volume}{11}},
  \bibinfo{pages}{850--854} (\bibinfo{year}{2012}).

\bibitem{samal_manipulating_2015}
\bibinfo{author}{Samal, D.} \& \bibinfo{author}{Koster, G.}
\newblock \bibinfo{title}{Manipulating oxygen sublattice in ultrathin cuprates:
  {A} new direction to engineer oxides}.
\newblock \emph{\bibinfo{journal}{Journal of Materials Research}}
  \textbf{\bibinfo{volume}{30}}, \bibinfo{pages}{463--476}
  (\bibinfo{year}{2015}).

\bibitem{Mannhart_Oxide_2010}
\bibinfo{author}{Mannhart, J.} \& \bibinfo{author}{Schlom, D.~G.}
\newblock \bibinfo{title}{Oxide {Interfaces}?{An} {Opportunity} for
  {Electronics}}.
\newblock \emph{\bibinfo{journal}{Science}} \textbf{\bibinfo{volume}{327}},
  \bibinfo{pages}{1607--1611} (\bibinfo{year}{2010}).

\bibitem{Zubko_Interface_2011}
\bibinfo{author}{Zubko, P.}, \bibinfo{author}{Gariglio, S.},
  \bibinfo{author}{Gabay, M.}, \bibinfo{author}{Ghosez, P.} \&
  \bibinfo{author}{Triscone, J.-M.}
\newblock \bibinfo{title}{Interface {Physics} in {Complex} {Oxide}
  {Heterostructures}}.
\newblock \emph{\bibinfo{journal}{Annual Review of Condensed Matter Physics}}
  \textbf{\bibinfo{volume}{2}}, \bibinfo{pages}{141--165}
  (\bibinfo{year}{2011}).

\bibitem{Zhong_Prediction_2012}
\bibinfo{author}{Zhong, Z.}, \bibinfo{author}{Koster, G.} \&
  \bibinfo{author}{Kelly, P.~J.}
\newblock \bibinfo{title}{Prediction of thickness limits of ideal polar
  ultrathin films}.
\newblock \emph{\bibinfo{journal}{Phys. Rev. B}} \textbf{\bibinfo{volume}{85}},
  \bibinfo{pages}{121411} (\bibinfo{year}{2012}).

\bibitem{Samal_Experimental_2013}
\bibinfo{author}{Samal, D.} \emph{et~al.}
\newblock \bibinfo{title}{Experimental {Evidence} for {Oxygen} {Sublattice}
  {Control} in {Polar} {Infinite} {Layer} {SrCuO}2}.
\newblock \emph{\bibinfo{journal}{Phys. Rev. Lett.}}
  \textbf{\bibinfo{volume}{111}}, \bibinfo{pages}{096102}
  (\bibinfo{year}{2013}).

\bibitem{Kuiper_Control_2013}
\bibinfo{author}{Kuiper, B.} \emph{et~al.}
\newblock \bibinfo{title}{Control of oxygen sublattice structure in ultra-thin
  {SrCuO}2 films studied by {X}-ray photoelectron diffraction}.
\newblock \emph{\bibinfo{journal}{APL Materials}} \textbf{\bibinfo{volume}{1}},
  \bibinfo{pages}{042113} (\bibinfo{year}{2013}).

\bibitem{le_tacon_intense_2011}
\bibinfo{author}{Le~Tacon, M.} \emph{et~al.}
\newblock \bibinfo{title}{Intense paramagnon excitations in a large family of
  high-temperature superconductors}.
\newblock \emph{\bibinfo{journal}{Nat Phys}} \textbf{\bibinfo{volume}{7}},
  \bibinfo{pages}{725--730} (\bibinfo{year}{2011}).

\bibitem{dean_persistence_2013}
\bibinfo{author}{Dean, M. P.~M.} \emph{et~al.}
\newblock \bibinfo{title}{Persistence of magnetic excitations in
  {La}2-{xSrxCuO}4 from the undoped insulator to the heavily overdoped
  non-superconducting metal}.
\newblock \emph{\bibinfo{journal}{Nat Mater}} \textbf{\bibinfo{volume}{advance
  online publication}} (\bibinfo{year}{2013}).

\bibitem{lipscombe_persistence_2007}
\bibinfo{author}{Lipscombe, O.~J.}, \bibinfo{author}{Hayden, S.~M.},
  \bibinfo{author}{Vignolle, B.}, \bibinfo{author}{McMorrow, D.~F.} \&
  \bibinfo{author}{Perring, T.~G.}
\newblock \bibinfo{title}{Persistence of high-frequency spin fluctuations in
  overdoped superconducting {La}2-{xSrxCuO}4 (x=0.22)}.
\newblock \emph{\bibinfo{journal}{Phys. Rev. Lett.}}
  \textbf{\bibinfo{volume}{99}}, \bibinfo{pages}{067002}
  (\bibinfo{year}{2007}).

\bibitem{Monney_Determining_2013}
\bibinfo{author}{Monney, C.} \emph{et~al.}
\newblock \bibinfo{title}{Determining the short-range spin correlations in the
  spin-chain {Li}2cuo2 and {CuGeO}3 compounds using resonant inelastic x-ray
  scattering}.
\newblock \emph{\bibinfo{journal}{Phys. Rev. Lett.}}
  \textbf{\bibinfo{volume}{110}}, \bibinfo{pages}{087403}
  (\bibinfo{year}{2013}).

\bibitem{Schlappa_Collective_2009}
\bibinfo{author}{Schlappa, J.} \emph{et~al.}
\newblock \bibinfo{title}{Collective {Magnetic} {Excitations} in the {Spin}
  {Ladder} {Sr}14cu24o41 {Measured} {Using} {High}-{Resolution} {Resonant}
  {Inelastic} {X}-{Ray} {Scattering}}.
\newblock \emph{\bibinfo{journal}{Physical Review Letters}}
  \textbf{\bibinfo{volume}{103}} (\bibinfo{year}{2009}).

\bibitem{Lee_Role_2013}
\bibinfo{author}{Lee, W.~S.} \emph{et~al.}
\newblock \bibinfo{title}{Role of {Lattice} {Coupling} in {Establishing}
  {Electronic} and {Magnetic} {Properties} in {Quasi}-{One}-{Dimensional}
  {Cuprates}}.
\newblock \emph{\bibinfo{journal}{Phys. Rev. Lett.}}
  \textbf{\bibinfo{volume}{110}}, \bibinfo{pages}{265502}
  (\bibinfo{year}{2013}).

\bibitem{Schlappa_Spin-orbital_2012}
\bibinfo{author}{Schlappa, J.} \emph{et~al.}
\newblock \bibinfo{title}{Spin-orbital separation in the quasi-one-dimensional
  {Mott} insulator {Sr}2cuo3}.
\newblock \emph{\bibinfo{journal}{Nature}} \textbf{\bibinfo{volume}{485}},
  \bibinfo{pages}{82--85} (\bibinfo{year}{2012}).

\bibitem{Sala_Energy_2011}
\bibinfo{author}{Sala, M.~M.} \emph{et~al.}
\newblock \bibinfo{title}{Energy and symmetry of dd excitations in undoped
  layered cuprates measured by {Cu} {L}3 resonant inelastic x-ray scattering}.
\newblock \emph{\bibinfo{journal}{New J. Phys.}} \textbf{\bibinfo{volume}{13}},
  \bibinfo{pages}{043026} (\bibinfo{year}{2011}).

\bibitem{Bisogni_Orbital_2015}
\bibinfo{author}{Bisogni, V.} \emph{et~al.}
\newblock \bibinfo{title}{Orbital {Control} of {Effective} {Dimensionality}:
  {From} {Spin}-{Orbital} {Fractionalization} to {Confinement} in the
  {Anisotropic} {Ladder} {System} {CaCu}2o3}.
\newblock \emph{\bibinfo{journal}{Phys. Rev. Lett.}}
  \textbf{\bibinfo{volume}{114}}, \bibinfo{pages}{096402}
  (\bibinfo{year}{2015}).

\bibitem{Dean_High-Energy_2013}
\bibinfo{author}{Dean, M. P.~M.} \emph{et~al.}
\newblock \bibinfo{title}{High-{Energy} {Magnetic} {Excitations} in the
  {Cuprate} {Superconductor} {Bi}2sr2cacu2o8+delta: {Towards} a {Unified}
  {Description} of {Its} {Electronic} and {Magnetic} {Degrees} of {Freedom}}.
\newblock \emph{\bibinfo{journal}{Phys. Rev. Lett.}}
  \textbf{\bibinfo{volume}{110}}, \bibinfo{pages}{147001}
  (\bibinfo{year}{2013}).

\bibitem{Zhou_Persistent_2013}
\bibinfo{author}{Zhou, K.-J.} \emph{et~al.}
\newblock \bibinfo{title}{Persistent high-energy spin excitations in
  iron-pnictide superconductors}.
\newblock \emph{\bibinfo{journal}{Nat Commun}} \textbf{\bibinfo{volume}{4}},
  \bibinfo{pages}{1470} (\bibinfo{year}{2013}).

\bibitem{Guarise_Anisotropic_2014}
\bibinfo{author}{Guarise, M.} \emph{et~al.}
\newblock \bibinfo{title}{Anisotropic softening of magnetic excitations along
  the nodal direction in superconducting cuprates}.
\newblock \emph{\bibinfo{journal}{Nat Commun}} \textbf{\bibinfo{volume}{5}}
  (\bibinfo{year}{2014}).

\bibitem{strocov_high-resolution_2010}
\bibinfo{author}{Strocov, V.~N.} \emph{et~al.}
\newblock \bibinfo{title}{High-resolution soft {X}-ray beamline {ADRESS} at the
  {Swiss} {Light} {Source} for resonant inelastic {X}-ray scattering and
  angle-resolved photoelectron spectroscopies}.
\newblock \emph{\bibinfo{journal}{J Synchrotron Radiat}}
  \textbf{\bibinfo{volume}{17}}, \bibinfo{pages}{631--643}
  (\bibinfo{year}{2010}).

\bibitem{Ghiringhelli_SAXES_2006}
\bibinfo{author}{Ghiringhelli, G.} \emph{et~al.}
\newblock \bibinfo{title}{{SAXES}, a high resolution spectrometer for resonant
  x-ray emission in the 400?1600ev energy range}.
\newblock \emph{\bibinfo{journal}{Review of Scientific Instruments}}
  \textbf{\bibinfo{volume}{77}}, \bibinfo{pages}{113108}
  (\bibinfo{year}{2006}).

\bibitem{Zaliznyak_Spinons_2004}
\bibinfo{author}{Zaliznyak, I.~A.} \emph{et~al.}
\newblock \bibinfo{title}{Spinons in the {Strongly} {Correlated} {Copper}
  {Oxide} {Chains} in {SrCuuO}2}.
\newblock \emph{\bibinfo{journal}{Phys. Rev. Lett.}}
  \textbf{\bibinfo{volume}{93}}, \bibinfo{pages}{087202}
  (\bibinfo{year}{2004}).

\bibitem{Walters_Effect_2009}
\bibinfo{author}{Walters, A.~C.} \emph{et~al.}
\newblock \bibinfo{title}{Effect of covalent bonding on magnetism and the
  missing neutron intensity in copper oxide compounds}.
\newblock \emph{\bibinfo{journal}{Nat Phys}} \textbf{\bibinfo{volume}{5}},
  \bibinfo{pages}{867--872} (\bibinfo{year}{2009}).

\bibitem{takano_acuo2_1989}
\bibinfo{author}{Takano, M.}, \bibinfo{author}{Takeda, Y.},
  \bibinfo{author}{Okada, H.}, \bibinfo{author}{Miyamoto, M.} \&
  \bibinfo{author}{Kusaka, T.}
\newblock \bibinfo{title}{{ACuO}2 ({A}: alkaline earth) crystallizing in a
  layered structure}.
\newblock \emph{\bibinfo{journal}{Physica C: Superconductivity}}
  \textbf{\bibinfo{volume}{159}}, \bibinfo{pages}{375--378}
  (\bibinfo{year}{1989}).

\bibitem{Hyatt_High-pressure_2004}
\bibinfo{author}{Hyatt, N.~C.}, \bibinfo{author}{Gray, L.},
  \bibinfo{author}{Gameson, I.}, \bibinfo{author}{Edwards, P.~P.} \&
  \bibinfo{author}{Hull, S.}
\newblock \bibinfo{title}{High-pressure neutron diffraction study of the
  quasi-one-dimensional cuprate {Sr}2cuo3}.
\newblock \emph{\bibinfo{journal}{Phys. Rev. B}} \textbf{\bibinfo{volume}{70}},
  \bibinfo{pages}{214101} (\bibinfo{year}{2004}).

\bibitem{Minola_Magnetic_2012}
\bibinfo{author}{Minola, M.} \emph{et~al.}
\newblock \bibinfo{title}{Magnetic and ligand field properties of copper at the
  interfaces of ({CaCuO}2)n/({SrTiO}3)n superlattices}.
\newblock \emph{\bibinfo{journal}{Phys. Rev. B}} \textbf{\bibinfo{volume}{85}},
  \bibinfo{pages}{235138} (\bibinfo{year}{2012}).

\bibitem{Ament_Theoretical_2009}
\bibinfo{author}{Ament, L. J.~P.}, \bibinfo{author}{Ghiringhelli, G.},
  \bibinfo{author}{Sala, M.~M.}, \bibinfo{author}{Braicovich, L.} \&
  \bibinfo{author}{van~den Brink, J.}
\newblock \bibinfo{title}{Theoretical {Demonstration} of {How} the {Dispersion}
  of {Magnetic} {Excitations} in {Cuprate} {Compounds} can be {Determined}
  {Using} {Resonant} {Inelastic} {X}-{Ray} {Scattering}}.
\newblock \emph{\bibinfo{journal}{Phys. Rev. Lett.}}
  \textbf{\bibinfo{volume}{103}}, \bibinfo{pages}{117003}
  (\bibinfo{year}{2009}).

\bibitem{braicovich_magnetic_2010}
\bibinfo{author}{Braicovich, L.} \emph{et~al.}
\newblock \bibinfo{title}{Magnetic {Excitations} and {Phase} {Separation} in
  the {Underdoped} {La}2-{xSrCuO}4 {Superconductor} {Measured} by {Resonant}
  {Inelastic} {X}-{Ray} {Scattering}}.
\newblock \emph{\bibinfo{journal}{Physical Review Letters}}
  \textbf{\bibinfo{volume}{104}} (\bibinfo{year}{2010}).

\bibitem{braicovich_dispersion_2009}
\bibinfo{author}{Braicovich, L.} \emph{et~al.}
\newblock \bibinfo{title}{Dispersion of {Magnetic} {Excitations} in the
  {Cuprate} {La}2cuo4 and {CaCuO}2 {Compounds} {Measured} {Using} {Resonant}
  {X}-{Ray} {Scattering}}.
\newblock \emph{\bibinfo{journal}{Phys. Rev. Lett.}}
  \textbf{\bibinfo{volume}{102}}, \bibinfo{pages}{167401}
  (\bibinfo{year}{2009}).

\bibitem{Coldea_Spin_2001}
\bibinfo{author}{Coldea, R.} \emph{et~al.}
\newblock \bibinfo{title}{Spin waves and electronic interactions in {La}2cuo4}
  (\bibinfo{year}{2001}).

\bibitem{headings_anomalous_2010}
\bibinfo{author}{Headings, N.~S.}, \bibinfo{author}{Hayden, S.~M.},
  \bibinfo{author}{Coldea, R.} \& \bibinfo{author}{Perring, T.~G.}
\newblock \bibinfo{title}{Anomalous {High}-{Energy} {Spin} {Excitations} in the
  {High}-\$\{{T}\}\_\{c\}\$ {Superconductor}-{Parent} {Antiferromagnet}
  {La}2cuo4}.
\newblock \emph{\bibinfo{journal}{Phys. Rev. Lett.}}
  \textbf{\bibinfo{volume}{105}}, \bibinfo{pages}{247001}
  (\bibinfo{year}{2010}).

\bibitem{Zhang_Effective_1988}
\bibinfo{author}{Zhang, F.} \& \bibinfo{author}{Rice, T.}
\newblock \bibinfo{title}{Effective {Hamiltonian} for the superconducting {Cu}
  oxides}.
\newblock \emph{\bibinfo{journal}{Physical Review B}}
  \textbf{\bibinfo{volume}{37}}, \bibinfo{pages}{3759--3761}
  (\bibinfo{year}{1988}).

\bibitem{Armitage_Progress_2010}
\bibinfo{author}{Armitage, N.~P.}, \bibinfo{author}{Fournier, P.} \&
  \bibinfo{author}{Greene, R.~L.}
\newblock \bibinfo{title}{Progress and perspectives on electron-doped
  cuprates}.
\newblock \emph{\bibinfo{journal}{Rev. Mod. Phys.}}
  \textbf{\bibinfo{volume}{82}}, \bibinfo{pages}{2421--2487}
  (\bibinfo{year}{2010}).

\bibitem{Lee_Asymmetry_2014}
\bibinfo{author}{Lee, W.~S.} \emph{et~al.}
\newblock \bibinfo{title}{Asymmetry of collective excitations in electron- and
  hole-doped cuprate superconductors}.
\newblock \emph{\bibinfo{journal}{Nat Phys}} \textbf{\bibinfo{volume}{10}},
  \bibinfo{pages}{883--889} (\bibinfo{year}{2014}).

\bibitem{Ishii_High-Energy_2014}
\bibinfo{author}{Ishii, K.} \emph{et~al.}
\newblock \bibinfo{title}{High-energy spin and charge excitations in
  electron-doped copper oxide superconductors}.
\newblock \emph{\bibinfo{journal}{Nat Commun}} \textbf{\bibinfo{volume}{5}}
  (\bibinfo{year}{2014}).

\end{thebibliography}
\end{document}